\documentstyle[eqsecnum,aps,epsfig]{revtex}
\begin{document}
\title{CONSTRUCTING THE FERMION-BOSON VERTEX IN QED3}
\author{A. Bashir and A. Raya} 
\address{Instituto de F{\'\i}sica y Matem\'aticas, 
         Universidad Michoacana de San Nicol\'as de Hidalgo\\
         Apartado Postal 2-82, Morelia, Michoac\'an 58040, M\'exico.}
\maketitle
\begin{abstract} 
We derive perturbative constraints on the transverse part of the 
fermion-boson vertex in massive QED3 through its one loop evaluation
in an arbitrary covariant gauge. Written in a particular form, these 
constraints naturally  lead us to the
first non-perturbative construction of the vertex, which is in complete
agreement with its one loop expansion in all momentum regimes.
Without affecting its one-loop perturbative properties,
we also construct an effective vertex in such a way that
the unknown functions defining it have no dependence on the angle between
the incoming and outgoing fermion momenta. Such a vertex should be useful
for the numerical study of dynamical chiral symmetry breaking, 
leading to more reliable results.
\end{abstract}

\vspace{4mm}
\hspace{1.6cm}PACS numbers: 12.20.Ds, 11.15Tk  
\hspace{2.5cm} Preprint number: UMSNH-PHYS/01-6

\section{Introduction}

     Quantum Electrodynamics in 3-dimensions (QED3) is an attractive model 
to study the intricacies of Schwinger-Dyson Equations (SDEs). Due to its 
simplicity as compared to Quantum Electrodynamics in 4-dimensions (QED4) 
and Quantum Chromodynamics (QCD), the corresponding study of dynamical 
symmetry breaking is relatively neater in QED3. There exist numerous works 
both in the quenched and the unquenched approximation in this connection, 
e.g., \cite{quenched1}-\cite{unquenched7}.  An excellent review 
can be found in \cite{review}. As is well known, the knowledge of the 
3-point vertex is crucial in such studies. In this respect, perturbation
theory is a powerful point of reference as it is natural to believe
that physically meaningful solutions of the Schwinger-Dyson equations must 
agree with perturbative results in the weak coupling regime. This realization
has been exploited in \cite{BKP1,BKP2,AB4,AB5} to derive constraints on 
the fermion propagator and the three-point vertex in massless QED3. In this 
paper, we extend this work to the more general massive case, based on 
\cite{raya1}.

     The Ward-Takahashi Identity (WTI) relates the 3-point vertex to the 
fermion propagator. Using this relation, a part of the vertex, called 
longitudinal, can be expressed in terms of the fermion propagator ~\cite{BC}.
We evaluate this propagator to one-loop and hence determine the
longitudinal vertex to the same order. We also calculate the complete
vertex to one-loop and a mere subtraction of the longitudinal part
yields the transverse part, the one which is not fixed by the WTI.
According to the choice of Ball and Chiu, which was later modified by
K{\i}z{\i}lers\"{u} {\it et. al.} \cite{KRP}, the transverse vertex
can be expressed in terms of 8 independent spin structures. 
The vertex should be free of any kinematic singularities.
Ball and Chiu chose the basis in such a way that the coefficient of each
of the basis is independently free of kinematic singularities
in the Feynman gauge. It was later shown by K{\i}z{\i}lers\"{u} {\it et. al.} 
\cite{KRP} that a calculation similar to that of Ball and Chiu 
in an arbitrary covariant gauge does not have the same nice feature.
Therefore, they proposed a modified basis whose coefficients are
free of kinematic singularities in an arbitrary covariant gauge.
The calculation in the present  paper confirms that all the vectors
of the modified basis also retain this feature for massive QED3. The final 
result for the transverse vertex is written in terms of basic functions of 
the momenta in a form suitable for its extension to the non-perturbative 
domain, following the ideas of Curtis and Pennington \cite{Davy}.

       Using perturbative constraints as a guide, we carry out a 
construction of the non-perturbative vertex, which has no explicit 
dependence on the coupling $\alpha$. This vertex has an explicit dependence 
on the gauge parameter $\xi$. We demonstrate in the massless case that a 
vertex cannot be constructed without an explicit dependence on $\xi$. 
For practical purposes of the numerical study of dynamical chiral
symmetry breaking, we also construct an effective 
vertex which shifts the angular dependence from the unknown fermion
propagator functions to the known basic functions, without changing its
perturbative properties at the one-loop level. We believe that this
vertex should lead to a more realistic study of the dynamically generated
masses through the corresponding SDEs. 

\newpage

\section{Longitudinal and Transverse Vertex to One-Loop}
\subsection{The Fermion Propagator}
\vspace{5mm}

    One-loop fermion propagator 
can be obtained by evaluating the graph in Fig.~1. This graph corresponds
to the following equation~:
\begin{eqnarray}
  i S_F^{-1}(p) &=& i {S_F^0}^{-1}(p) + e^2 \; \int \frac{d^3k}{(2 \pi)^3}
  \; \gamma^{\mu} \,  S_F^0(k) \, \gamma^{\nu} \, \Delta_{\mu \nu}^0(q)
\label{propSDE}   \;,
\end{eqnarray}
where $q=k-p$ and $e$ is the QED coupling constant. The bare fermion and
photon propagators are, respectively~: 
\begin{eqnarray}
      {S_F^0}(p) &=&  \frac{1}{{\not\! p}-m}  \;,  \nonumber \\
   \Delta^0_{\mu\nu}(q)&=&- \left[q^2
 g_{\mu\nu}+(\xi-1)q_{\mu}q_{\nu}\right]/q^4 \;, \label{bareprop}
\end{eqnarray}
where $m$ is the bare mass of the fermion and $\xi$ is the covariant
gauge parameter. 
We define the full fermion propagator $S_{F}(p)$ in the most general form as~:
\begin{eqnarray}
      S_{F}(p) &=&    \frac{F(p^2)}{{\not\! p}-{\cal M}(p^2)}  \;. 
\label{fullprop}
\end{eqnarray}
     Taking the trace of Eq.~(\ref{propSDE}), having multiplied it with 
$\not\! p$ and  with $1$ respectively, one can obtain two independent 
equations. On simplifying, these equations can be written as~:
\begin{eqnarray}
  \frac{1}{F(p^2)} &=& 1 \; + i 4 \pi \alpha \xi \, \frac{1}{p^2}
  \int \frac{d^3k}{(2 \pi)^3}  \;\frac{1}{q^4(k^2-m^2)} \;    
\left[ (k^2+p^2) k \cdot p - 2 k^2 p^2 \right] \; , \label{Fangular}
\\ \nonumber \\
 \frac{{\cal M}(p^2)}{F(p^2)} &=& m 
-i 4\pi\alpha \, (\xi+2) \int \frac{d^3k}{(2 \pi)^3} \; 
\frac{m}{q^2(k^2-m^2)}  \;,   \label{Mangular}
\end{eqnarray}
where $\alpha=e^2/4 \pi$.
On Wick rotating to the Euclidean space and carrying out angular and 
radial integrations,
we arrive at~:
\begin{eqnarray}
\frac{1}{F(p^2)}&=&1-\frac{\alpha\xi}{2 p^2}  \, 
\left[ m - (m^2+p^2) \, I(p^2) 
\right] \;,\nonumber\\ \nonumber \\
\frac{{\cal M}(p^2)}{F(p^2)}&=&m\left[1+
\alpha (\xi+2) \, I(p^2) \right] \;, \label{FMradial}
\end{eqnarray}
where we have used the simplifying notation $I(p^2)=  
({1}/{\sqrt{-p^2}}) \arctan \sqrt{{-p^{2}}/{m^{2}}}$.
Equations~(\ref{fullprop}) and~(\ref{FMradial}) form the complete
fermion propagator at one loop.

\subsection{Longitudinal Vertex to One Loop}

The full vertex satisfies WTI
\begin{eqnarray}
q_{\mu}\Gamma^{\mu}(k,p)={\it S}^{-1}_{F}(k)-{\it S}^{-1}_{F}(p)  \; .
\label{WTI}
\end{eqnarray}
This relation allows us to decompose the full vertex into 
longitudinal $( \Gamma^{\mu}_{L}(k,p) )$ and 
transverse $( \Gamma^{\mu}_{T}(k,p) )$ parts~:
\begin{eqnarray}
\Gamma^{\mu}(k,p)=\Gamma^{\mu}_{L}(k,p)+\Gamma^{\mu}_{T}(k,p) \;,
\label{vertex}
\end{eqnarray}
where the transverse part satisfies
\begin{eqnarray}
q_{\mu}\Gamma^{\mu}_{T}(k,p)=0\;\;\;\;\;\mbox{and} \;\;\;\;
\Gamma^{\mu}_{T}(p,p)=0 \label{defofTvertex}
\end{eqnarray}
and  hence remains  undetermined  by WTI. Following the work of Ball and 
Chiu, we can define the longitudinal component of the vertex in terms of
the fermion propagator as
\begin{eqnarray}
\Gamma^{\mu}_{L}&=&\frac{\gamma^{\mu}}{2}
\left[ \frac{1}{F(k^2)}+\frac{1}{F(p^2)} \right] \; + \; 
\frac{1}{2} \, \frac{({\not \! k}+{\not \! p})(k+p)^{\mu}}
{(k^2-p^2)}\left[ \frac{1}{F(k^2)}-\frac{1}{F(p^2)} \right] \; 
+ \;  \frac{(k+p)^{\mu}}
{(k^2-p^2)}\left[ \frac{{\cal M}(k^2)}{F(k^2)}-
\frac{{\cal M}(p^2)}{F(p^2)}\right] \; . \label{Lvertex}
\end{eqnarray}
On substituting Eq.~(\ref{FMradial}) into the above expression, we obtain   
\begin{eqnarray}
    \Gamma^{\mu}_{L} = \left[ 1 + \frac{\alpha \xi}{4} \, \sigma_1 \right]
    \, \gamma^{\mu} \; + \;  \frac{\alpha \xi}{4} \, \sigma_2 \, \left[ 
{k^{\mu}}{\not \! k} \, +  \, {p^{\mu}}{\not \! p} \, + \,
{k^{\mu}}{\not \! p} \, +  \, {p^{\mu}}{\not \! k}  \right] \; + \;
\alpha (\xi+2) \sigma_3 \, \left[ k^{\mu} + p^{\mu}  \right],
\label{1loopLvertex}
\end{eqnarray}
where
\begin{eqnarray}
\sigma_1 &=&   \frac{m^2+k^2}{k^2}  \, I(k^2) \; + \; 
 \frac{m^2+p^2}{p^2}  \, I(p^2) \; - \; m \frac{k^2+p^2}{k^2 p^2} \;, 
  \nonumber \\ \nonumber \\
\sigma_2 &=& \frac{1}{k^2-p^2} \, \left[ 
 \frac{m^2+k^2}{k^2}  \, I(k^2) \; - \; 
 \frac{m^2+p^2}{p^2}  \, I(p^2) \; + \; m \frac{k^2-p^2}{k^2 p^2}
  \right] \;,  \nonumber  \\ \nonumber \\
\sigma_3 &=& m \;  \left[ I(k^2) \; - \; I(p^2)  \right]   \;.
\label{sigmas}
\end{eqnarray}
Eqs.~(\ref{1loopLvertex}) and~(\ref{sigmas}) give the longitudinal part of the 
fermion-photon vertex to one loop for the massive QED3.

\subsection{Transverse Vertex to One-Loop}

  The vertex of Fig.~2 can be expressed as
\begin{eqnarray}
\Gamma^{\mu}(k,p)=\,\gamma^{\mu}+\,\Lambda^{\mu} \;.  \label{01loopvertex}
\end{eqnarray}
Using the Feynman rules, ${\Lambda^{\mu}}$ to
${O(\alpha)}$ is simply given by~:
\begin{eqnarray}
-{\it i}e\Lambda^{\mu}\,=\,\int_{M}\frac{d^3w}{(2\,\pi)^3}
(-{\it i}e\gamma^{\alpha}){\it i}{\it S}^{0}_{F}(p-w)(-{\it i}e\gamma^{\mu})
{\it i}{\it S}^{0}_{F}(k-w)(-{\it i}e\gamma^{\beta}){\it i}
\Delta^0_{\alpha\beta}(w) \;,  \label{1loopvertex}
\end{eqnarray}
where the loop integral is to be performed in Minkowski space. $\Lambda^{\mu}$
can be expressed as~:
\begin{eqnarray}
\Lambda^{\mu}&=&-\frac{{\it i}\,{\alpha}}{2\,{\pi}^2}
\Bigg\{ \left[ \gamma^{\alpha} {\not \! p}\,{\gamma^{\mu}}\,{\not \! k} 
 \gamma_{\alpha}   + m ( 4 k^{\mu} + 4 p^{\mu} -  {\not \! p} 
{\gamma^{\mu}} -  {\gamma^{\mu}}   {\not \! k} ) -m^2
{\gamma^{\mu}}       \right] 
{\it J}^{(0)}  \nonumber \\
&&\hspace{10mm} -  \left[ {\gamma^{\alpha}}
 {\not \! p}\,{\gamma^{\mu}}{\gamma^{\nu}} \gamma_{\alpha}
+ \gamma^{\alpha}  {\gamma^{\nu}} {\gamma^{\mu}} {\not \! k} \gamma_{\alpha}
+ 6 m g^{\mu \nu}     \right] 
{{\it J}_{\nu}^{(1)}}
+{\gamma^{\alpha}}{\gamma^{\nu}}{\gamma^{\mu}}{\gamma^{\lambda}}
\gamma_{\alpha}{{\it J}_{\nu\lambda}^{(2)}}\nonumber\\
&& \hspace{10mm}
+(\xi-1)\Bigg[ {\gamma^{\mu}}{\it K}^{(0)} - 
\left[ {\gamma^{\nu}}{\not \! p}\,{\gamma^{\mu}}+\,
{\gamma^{\mu}}{\not \! k}\,{\gamma^{\nu}}+2m g^{\mu \nu}
\right]
{\it J}_{\nu}^{(1)}
\nonumber\\
&& \hspace{10mm} +  \left[ {\gamma^{\nu}}{\not \! p}\,
{\gamma^{\mu}}{\not \! k}\,
{\gamma^{\lambda}}
+m(\,{\gamma^{\nu}}{\not \! p}\,{\gamma^{\mu}}\,{\gamma^{\lambda}}+
\,{\gamma^{\nu}}\,{\gamma^{\mu}}\,{\not \! k} \,{\gamma^{\lambda}})
+m^2\,{\gamma^{\nu}}\,{\gamma^{\mu}}\,{\gamma^{\lambda}}  \right] 
{{\it I}_{\nu\lambda}^{(2)}}
\Bigg]
\Bigg\}, \label{1loopvertexevaluated}
\end{eqnarray}
where the integrals
${K^{(0)}}$, $J^{(0)}$, $J^{(1)}_{\mu}$, $J^{(2)}_{\mu\nu}$, $I^{(0)}$, 
$I^{(1)}_{\mu}$ and  $I^{(2)}_{\mu\nu}$ are~:
\begin{eqnarray}
{\it K}^{(0)}&=&\int_{M}\,d^3w\,\frac{1}{[(p-w)^2-m^2]\,[(k-w)^2-m^2]} 
\nonumber \\
{\it J}^{(0)}&=&\int_{M}\,d^3w\,\frac{1}{w^2\,[(p-w)^2-m^2]\,[(k-w)^2-m^2]}
\nonumber \\
{\it J}^{(1)}_{\mu}&=&\int_{M}\,d^3w\,
\frac{w_{\mu}}{w^2\,[(p-w)^2-m^2]\,[(k-w)^2-m^2]}
\nonumber \\
{\it J}^{(2)}_{\mu\nu}&=&\int_{M}\,d^3w\,
\frac{w_{\mu}w_{\nu}}{w^2\,[(p-w)^2-m^2]\,[(k-w)^2-m^2]}
\nonumber \\
{\it I}^{(0)}&=&\int_{M}\,d^3w\,
\frac{1}{w^4\,[(p-w)^2-m^2]\,[(k-w)^2-m^2]}
\nonumber \\
{\it I}^{(1)}_{\mu}&=&\int_{M}\,d^3w\,
\frac{w_{\mu}}{w^4\,[(p-w)^2-m^2]\,[(k-w)^2-m^2]}
\nonumber \\
{\it I}^{(2)}_{\mu\nu}&=&\int_{M}\,d^3w\,
\frac{w_{\mu}w_{\nu}}{w^4\,[(p-w)^2-m^2]\,[(k-w)^2-m^2]}
\label{integrals}  \;.
\end{eqnarray}
We evaluate these integrals  following the techniques developed in
~\cite{BKP1,BKP2,BC,KRP}. The results are tabulated in the appendix, 
employing the notation $\Delta^2 = (k \cdot p)^2 - k^2 p^2$ and 
$X_0=(2/i\pi^2)X^{(0)}$ for $X=I,J,K$.
Having calculated the vertex to ${O(\alpha)}$, 
Eq.~(\ref{1loopvertexevaluated}), 
we can subtract from it the longitudinal vertex, 
Eqs.~(\ref{1loopLvertex}) and~(\ref{sigmas}), and obtain the transverse 
vertex to ${O(\alpha)}$. Following the scheme provided by Ball and Chiu 
\cite{BC}, and modified 
later by K{\i}z{\i}lers\"{u} {\it et. al.} \cite{KRP}, the transverse 
vertex ${\Gamma^{\mu}_{T}(k,p)}$ can be written in terms of 8 
basis vectors as follows~:
\begin{eqnarray}
\Gamma^{\mu}_{T}(k,p)=\sum_{i=1}^{8} \tau_{i}(k^2,p^2,q^2)T^{\mu}_{i}(k,p) 
\;,
\end{eqnarray}
where
\begin{eqnarray}
&T^{\mu}_{1}&=\left[p^{\mu}(k\cdot q)-k^{\mu}(p\cdot q)\right] \nonumber\\
&T^{\mu}_{2}&=\left[p^{\mu}(k\cdot q)-k^{\mu}(p\cdot q)\right]({\not\! k}
+{\not\! p})\nonumber\\
&T^{\mu}_{3}&=q^2\gamma^{\mu}-q^{\mu}{\not \! q}\nonumber\\
&T^{\mu}_{4}&=q^2\left[\gamma^{\mu}({\not \! k}+{\not \! p})-k^{\mu}
-p^{\mu}\right]-2(k-p)^{\mu}k^{\lambda}p^{\nu}\sigma_{\lambda\nu}\nonumber\\
&T^{\mu}_{5}&=q_{\nu}\sigma^{\nu\mu}\nonumber\\
&T^{\mu}_{6}&=-\gamma^{\mu}(k^2-p^2)+(k+p)^{\mu}{\not \! q}\nonumber\\
&T^{\mu}_{7}&=-\frac{1}{2}(k^2-p^2)\left[\gamma^{\mu}({\not \! k}+{\not \! p})
-k^{\mu}-p^{\mu} \right]
+(k+p)^{\mu}k^{\lambda}p^{\nu}\sigma_{\lambda\nu}\nonumber\\
&T^{\mu}_{8}&=-\gamma^{\mu}k^{\nu}p^{\lambda}{\sigma_{\nu\lambda}}
+k^{\mu}{\not \! p}-p^{\mu}{\not \! k} \;,\nonumber\\
\mbox{with}\;\;\;\;\;\;\;\;\;
&\sigma_{\mu\nu}&=\frac{1}{2}[\gamma_{\mu},\gamma_{\nu}]\;.
\end{eqnarray}
   After a lengthy but straightforward algebra, the coefficients 
${\tau_{\it i}}$ can be identified. We prefer to write these out in the 
following form~:
\begin{eqnarray}
    \tau_i(k,p) &=& \alpha g_i  \left[ \sum_j^5 a_{ij}(k,p) I(l_j^2) 
+ \frac{a_{i6}(k,p)}{k^2 p^2}  \right]  \hspace{10 mm} i=1,2, \cdots 8  \;,
\label{taui}
\end{eqnarray}
where $l_1^2 = \eta_1^2\chi/4$,
$l_2^2 = \eta_2^2\chi/4$,
$l_3^2 = k^2$,
$l_4^2 = p^2$ and 
$l_5^2 = q^2/4$. Functions $\eta_1$, $\eta_2$ and $\chi$ have been defined
in the appendix, Eqs.~(\ref{etaandchi}). Similarly, the factors $g_i$ are
$-g_1 = m \Delta^2 g_2 = 2m \Delta^2 g_3 = 2 \Delta^2 g_4 = 
g_5 = 2 m \Delta^2 g_6 = \Delta^2 g_7 = m g_8 = m/ 4 \Delta^2$.
The coefficients $a_{ij}(k,p)$ have also been tabulated in the appendix,
Eqs.~(\ref{coeftau}). An important point to note is that these coefficients
do not contain any trigonometric function, as it has been extracted out
for raising the $\tau_i$ to a non-perturbative status.
The ${\tau_i}$ have the required symmetry under 
the exchange of vectors $k$ and $p$. All the $\tau_i$ are symmetric except 
$\tau_4$ and ${\tau_6}$ which are antisymmetric. Note that the form in which 
we write the transverse vertex makes it clear that each term in all the 
$\tau_i$ is either proportional to $\alpha I(l^2)$ or $\alpha/(k^2 p^2)$. 
We shall see that this form provides us with a natural scheme to
arrive at its simple non-perturbative extension. 

A few comments in comparison with the work by
Davydychev {\em et. al.} \cite{Davy}, are as follows: (i)  None of the 
${\tau_i}$ we have calculated  has kinematic singularity when $k^2 \to p^2$.
This clearly suggests that the choice of the $\tau_i$ suggested by
K{\i}z{\i}lers\"{u} {\it et. al.} is preferred over the one of
Ball and Chiu (in QED3 as well) used by Davydychev {\em et. al.} 
\cite{Davy}. In particular our $\tau_4$ and $\tau_7$ are independent 
of kinematic singularities.
(ii) In three
dimensions, their factorization of the common constant factor in
Eq.~(E.1) is singular. However, as the divergences completely cancel
out, we find our expressions more suitable for writing the
transverse vertex in three dimensions. (iii) With the way we express
$J_0$, all the $\tau_i$ are written in terms of basic functions of
$k$ and $p$ and a single trigonometric function of the form $I(l^2)$.
This form plays a key role to enable us to make an easy transition to the 
possible non-perturbative structure of the vertex, as explained in the 
next section. Moreover, with the given form of $J_0$, a direct comparison 
can be made with the massless case.

\section{Non-perturbative form of the vertex}

\subsection{On the Gauge Parameter Dependence of the Vertex}

Let us first look at the $\tau_i$ in the simplified massless case, with the 
notation  $k= \sqrt{-k^2}$, $p=\sqrt{-p^2}$ and $q=\sqrt{-q^2}$
\cite{BKP1,BKP2},
\begin{eqnarray}
  \tau_2 &=& \frac{\alpha \pi}{4} \; \frac{1}{kp(k+p)(k+p+q)^2} \;
\left[  1 + (\xi-1) \, \frac{2k+2p+q}{q}  \right] , \\ \nonumber \\
   \tau_3 &=& \frac{\alpha \pi}{8} \; \frac{1}{kpq(k+p+q)^2} \;
 \left[ 4kp+3kq+3pq+2q^2 + (\xi-1) \, (2k^2+2p^2+kq+pq)  \right] , 
\nonumber\\
   \\
 \tau_6 &=& \frac{\alpha \pi (2- \xi)}{8} \; \frac{k-p}{kp(k+p+q)^2} \;,
\label{masslesstau6} \\ 
   \nonumber \\
 \tau_8 &=& \frac{\alpha \pi (2+ \xi)}{2} \; \frac{1}{kp(k+p+q)} \;.
\end{eqnarray}
It is interesting to note that the existence of
the factor 
\begin{eqnarray*}
    \frac{k-p}{kp} = -\left( \frac{1}{k} \; - \; \frac{1}{p} \right)
\end{eqnarray*}
in Eq.~(\ref{masslesstau6}) puts $\tau_6$ on a different footing as compared 
to the rest of the $\tau_i$. The reason is that in the massless limit, the 
fermion propagator is simply
\begin{eqnarray*}
   \frac{1}{F(p^2)} &=& 1 + \frac{\pi \alpha \xi}{4} \; \frac{1}{p} \;,
\end{eqnarray*}
implying 
\begin{eqnarray*}
 \frac{1}{F(k^2)} -  \frac{1}{F(p^2)} \propto 
\left[ \frac{1}{k} \; - \; \frac{1}{p} \right]
 \;.
\end{eqnarray*}
Therefore, the relation of $\tau_6$ with the fermion propagator of the type 
$[1/F(k^2)-1/F(p^2)]$ seems to arise rather
naturally~:
\begin{eqnarray}
  \tau_6 &=& - \frac{1}{2 \xi} \, \frac{2-\xi}{(k+p+q)^2} \;
\left[ \frac{1}{F(k^2)} - \frac{1}{F(p^2)} \right]\;,  \label{m0nptau6} 
\end{eqnarray}
as noticed first by Curtis and Pennington \cite{CP}. In the rest of the
$\tau_i$, the factor $1/k-1/p$ does not arise. However, one could 
introduce it by hand to arrive at the following expressions~:
\begin{eqnarray}
  \tau_2 &=& -\frac{1}{\xi} \; \frac{1}{(k^2-p^2)(k+p+q)^2} \;
\left(  1 + (\xi-1) \, \frac{2k+2p+q}{q}  \right) \;
\left[ \frac{1}{F(k^2)} - \frac{1}{F(p^2)} \right] 
 \\ \nonumber \\                                \label{m0nptau2}
   \tau_3 &=& -\frac{1}{2 \xi} \; \frac{1}{q(k-p)(k+p+q)^2} \;
 \left[ 4kp+3kq+3pq+2q^2 + (\xi-1) \, (2k^2+2p^2+kq+pq)  \right]  \;
\left[ \frac{1}{F(k^2)} - \frac{1}{F(p^2)} \right] 
 \label{m0nptau3}  \\
 \tau_8 &=& -\frac{2 (2+ \xi)}{\xi} \; \frac{1}{(k-p)(k+p+q)} 
\;\left[ \frac{1}{F(k^2)} - \frac{1}{F(p^2)} \right] \quad. \label{m0nptau8}
\end{eqnarray}
Eqs.~(\ref{m0nptau6}-\ref{m0nptau8}) represent a non-perturbative vertex 
which is in agreement with  its complete one-loop expansion. This vertex 
has been constructed in accordance with the form advocated, e.g., in 
\cite{quenched2,CP,AB1}. There are a couple of important points which need 
to be discussed here:
\begin{itemize}

\item

There
is an explicit dependence on the gauge parameter, $\xi$. A widespread belief 
has been that the gauge dependence of the vertex should solely arise through 
functions  $F(k^2)$ and $F(p^2)$, and there should
be {\em no explicit appearance of the gauge parameter $\xi$}. Such a belief
has been expressed (or is reflected) in various works to date, e.g.,   
\cite{quenched2,quenched3,unquenched7,CP,AB1}. Here we show that at least
in massless QED3, such a construction is not possible.

\item
One of the main reasons that the transverse vertex was believed to be 
proportional to the factor
\begin{eqnarray*}
\left[ \frac{1}{F(k^2)} - \frac{1}{F(p^2)} \right] 
\end{eqnarray*}
was the assumption that the transverse vertex vanishes in the Landau gauge.
This assumption was based upon the one-loop calculation of the vertex
in QED4 in the limit when the momentum in one of the fermion legs is much
greater as compared to the momentum in the other fermion leg \cite{CP}.
A complete one-loop calculation reveals that the transverse vertex
does not vanish in the Landau gauge. Moreover, an explicit presence of 
the gauge parameter in the non-perturbative form of the vertex 
tells us that the presence of the factor $ [1/F(k^2) - 1/F(p^2) ] $ is
no longer a guarantee that the transverse vertex vanishes in the 
Landau gauge.

\end{itemize}
We now show that the explicit dependence of the vertex on the gauge 
parameter $\xi$ is unavoidable in massless QED3.
We notice that at the one loop level, each of the $\tau_i$ can be written in 
the following form~:
\begin{eqnarray*}
\tau_{i}(k, p,q) &=& \alpha \xi \; a_{i}(k,p,q) + \alpha \; b_{i}(k,p,q) \;.
\end{eqnarray*}
On the other hand, Eq.~(\ref{FMradial}) yields the following form for $F$~:
\begin{eqnarray*}
     \frac{1}{F(p^2)} &=& 1 + \alpha \xi \; c_{i}(p) \;.
\end{eqnarray*}
If we want to write the non-perturbative form of the $\tau_{i}$ in terms of 
$1/F(p^2)$ and $1/F(k^2)$ 
alone and we do not expect explicit presence of $\alpha$, 
the only way to get rid of $\xi$  dependence is to have
\begin{eqnarray*}
b_{2} \; T_{2}^\mu + b_{3} \; T_{3}^\mu + b_{6} \; T_{6}^\mu + b_{8} \;
T_{8}^\mu = 0
\;.
\end{eqnarray*}
It is not possible as $T_i^{\mu}$ form a linearly independent set of basis
vectors. Therefore, any construction of the 3-point vertex will surely
have an explicit dependence on the gauge parameter.
 Owing to these reasons, we realize that to demand the transverse vertex 
to be proportional to $ [1/F(k^2) - 1/F(p^2) ] $ is artificial
(apart from $\tau_6$) and is not required. Therefore, we do not pursue this
line of action anymore. In the next section, we move on to construct the
vertex for the massive case inspired from our perturbative results.

\subsection{Non-perturbative Vertex}

      As pointed out in the previous section, each  term in all the 
$\tau_i$ is either proportional to the trigonometric function 
$\alpha I(l^2)$ or $\alpha/(k^2 p^2)$. 
On the other hand, the perturbative expressions for ${\cal M}(p^2)$ and
$F(p^2)$, Eqs.~(\ref{FMradial}), permit us to write~:
\begin{eqnarray}
\frac{1}{F(k^2)} - \frac{1}{F(p^2)} &=&\frac{\alpha}{k^2 p^2} \, \frac{\xi}{2}
\, \left[ k^2 \left\{ m- (m^2+p^2) I(p^2) \right\} - 
p^2 \left\{ m- (m^2+k^2) I(k^2) \right\}  \right]   \label{nonpF}
\end{eqnarray}
and
\begin{eqnarray}
\frac{\xi}{2 (2+\xi) l^2 I(l^2)} \, \left[ \frac{{\cal M}(l^2)}{F(l^2)}
- m \right] - \left[  1 - \frac{1}{F(l^2)}  \right] &=& 
\frac{\xi (m^2+l^2)}{2 l^2} \; \alpha I(l^2) \;.     \label{nonpM}
\end{eqnarray}
In the massless limit, Eq.~(\ref{nonpF}) simply reduces to 
\begin{eqnarray*}
\frac{1}{F(k^2)} - \frac{1}{F(p^2)} &=& \frac{\alpha \pi \xi}{4} \;
\left[  \frac{1}{k} - \frac{1}{p}  \right]    
\end{eqnarray*}
in the Euclidean space, as expected. It was in fact an analogous massless 
expression in the limit when $k >>p$ that inspired Curtis and Pennington, 
\cite{CP}, to propose their famous vertex in QED4. Here, we are extending 
the reasoning to all the momentum regimes in the massive QED3. Fortunate
simultaneouss occurrence of the factor $\alpha/(k^2 p^2)$ in all the 8
Eqs.~(\ref{taui}) and Eq.~(\ref{nonpF}), and the presence of the same
trigonometric factor $I(l^2)$ in the expressions for the vertex as well as
the propagator, one naturally arrives at the following non-perturbative form 
of $\tau_i$~:
\begin{eqnarray}
\nonumber
\tau_i &=& g_i\Bigg\{ \sum_{j=1}^5 \left( 
\frac{2a_{ij}(k,p)l_j^2}{\xi(m^2 + l_j^2)} \left[ 
\frac{\xi}{2  (\xi + 2) l_j^2  I(l_j^2)} \left(\frac{{\cal M}(l_j^2)}{F(l_j^2)}
-m \right) 
- \left( 1-\frac{1}{F(l_j^2)} \right) \right] \right)\\
&+& \frac{2 a_{i6}(k,p)}{\xi\left[k^2\left\{m-(m^2 + p^2)I(p^2)\right\}-
p^2\left\{m-(m^2 + k^2)I(k^2)\right\}\right]}
\left[ \frac{1}{F(k^2)}-\frac{1}{F(p^2)} \right] \Bigg\} \;.   
\label{npvertexangle}
\end{eqnarray}
By construction, in the weak coupling regime, this non-perturbative form of 
the transverse vertex reduces to its corresponding Feynman expansion at the
one loop level in an arbitrary covariant gauge and in all momentum regimes.
We would like to emphasize that this is not a unique non-perturbative 
construction. However, it is probably the most natural and the 
simplest. A two loop calculation similar to the one presented in our
paper, and the 
Landau-Khalatnikov transformation law for the vertex should serve as tests
of Eq.~(\ref{npvertexangle}) or guides for improvement towards the hunt for
the exact non-perturbative vertex. On practical side, the use of our
perturbation theory motivated vertex in studies addressing important issues 
such as dynamical mass generation for fundamental fermions should lead to 
more reliable results, attempting to preserve key features of gauge
field theories, e.g., gauge independence of physical observables.
A computational difficulty to use the above vertex  in such calculations 
could arise as the unknown functions $F$ and ${\cal M}$ depend on the 
angle between $k$ and $p$. This would make it impossible to carry out
angular integration analytically in the SDE for the fermion propagator. 
This problem can be circumvented 
by defining an effective vertex which shifts the angular dependence
from the unknown functions $F$ and ${\cal M}$ to the known basic functions
of $k$ and $p$. This can be done by re-writing the perturbative results,
Eq.~(\ref{taui}), as follows~:
\begin{eqnarray}
     \tau_i(k,p) &=& \alpha g_i \left[ b_{i1}(k,p) I(k^2) + b_{i2}(k,p) I(p^2)
   + \frac{a_{i6}(k,p)}{k^2 p^2}  \right]  \;,
\end{eqnarray}
where
\begin{eqnarray}
      b_{i1}(k,p)  &=& a_{i1}(k,p) \, \frac{I(l_1^2)}{I(l_3^2)} + a_{i3}(k,p) 
+ \frac{1}{2} \, a_{i5}(k,p) \frac{I(l_5^2)}{I(l_3^2)} \;,    \\
 b_{i2}(k,p)  &=& a_{i2}(k,p) \, \frac{I(l_2^2)}{I(l_4^2)} + a_{i4}(k,p) 
+ \frac{1}{2} \, a_{i5}(k,p) \frac{I(l_5^2)}{I(l_4^2)} \;.
\end{eqnarray}
This form can now be raised to a non-perturbative level exactly as before,
with the only difference that the functions $F$ and ${\cal M}$ are 
independent of the angle between the momenta $k$ and $p$~:
\begin{eqnarray}
\nonumber
\tau_i &=& g_i\Bigg\{ \sum_{j=1}^2 \left( 
\frac{2b_{ij}(k,p)\kappa_j^2}{\xi(m^2 + \kappa_{j}^2)} \left[ 
\frac{\xi}{2 (\xi + 2) \kappa_j^2 I(\kappa_j^2)  } 
\left(\frac{{\cal M}(\kappa_j^2)}{F(\kappa_j^2)}-m \right) 
- \left( 1-\frac{1}{F(\kappa_j^2)} \right) \right] \right)\\
&+& \frac{2  a_{i6}(k,p)}{\xi\left[k^2\left\{m-(m^2 + p^2)I(p^2)\right\}-
p^2\left\{m-(m^2 + k^2)I(k^2)\right\}\right]}
\left[ \frac{1}{F(k^2)}-\frac{1}{F(p^2)} \right] \Bigg\}  \;,
\end{eqnarray}
where $\kappa_1^2=k^2$ and $\kappa_2^2=p^2$. 

\section{Conclusions}

   In this paper, we calculate one loop fermion-boson vertex in QED3 in an 
arbitrary covariant gauge and write out the result in a form which naturally 
allows us to construct its non-perturbative counterpart. This is the first
construction of the  non-perturbative vertex which agrees with its Feynman
expansion in the weak coupling regime at the one loop level in all
momentum regimes and in arbitrary covariant gauge. For practical
numerical purposes, we also suggest a simple effective vertex
which shifts its angular dependence (angle between the incoming and outgoing
fermion momenta) from the fermion functions  to the known basic functions 
of the momenta involved, without affecting its perturbative properties
at the one-loop level. Currently, the work is underway to use this
vertex in numerical calculations of dynamical mass generation for the
fundamental fermions. We also plan to compare its gauge dependence against
the one  demanded by its Landau-Khalatnikov transformation~\cite{trans,LK}
in a non-perturbative fashion.

\section*{Acknowledgements}
  {}~~A.B. wishes to thank A. K{\i}z{\i}lers\"u with whom initial ideas
of the work were discussed. We are grateful to  A.I. Davydychev for answering 
a query. We acknowledge the CIC and the Conacyt grants under the projects 
4.12 and 32395-E, respectively.

\section*{Appendix}

Following are the results of the integrals listed in Eqs.~(\ref{integrals})~:

\noindent
\underline{\bf${\bf J^{(0)}}$}:
\begin{equation}
J_0 = \Bigg[ -\eta_1(k,p) I\left(\frac{{\eta_1}^2\chi}{4}\right) 
+ \eta_2(k,p) I\left(\frac{{\eta_2}^2\chi}{4}\right) \Bigg] \quad,
\label{J0}
\end{equation} 
with
\begin{eqnarray}
\eta_1(k,p) &=&-\Bigg\{ 
\frac{m^2(k^2-p^2)(2m^2- k^2 -p^2)+\chi}{\chi (m^2 - k^2)}\Bigg\} \;,  
                                                      \nonumber  \\
\eta_2(k,p) &=& -\eta_1(k,p)      \;,                 \nonumber  \\
\chi &=& m^2 (k^2-p^2)^2 \; + q^2(k^2-m^2)(p^2-m^2) \label{etaandchi} \;. 
\end{eqnarray}

\noindent
\underline{\bf${\bf K^{(0)}}$}:
\begin{eqnarray}
K^{(0)}&=& i \pi^2 \; I(q^2/4)  \;.
\end{eqnarray}

\noindent
\underline{\bf${\bf J^{(1)}_{\bf \mu}}$}:
\begin{eqnarray}
{\it J}^{(1)}_{\mu}=\frac{{\it i}\pi^2}{2}\left\{
k_{\mu}J_{A}(k,p)+p_{\mu}J_{B}(k,p)\right\} \;,
\end{eqnarray}
where
\begin{eqnarray}
J_{A}(k,p)&=&-\frac{2}{\Delta^2}   \Bigg\{  \left[ 
p^2 (k^2-k \cdot p) - m^2 (p^2-k \cdot p) \right] \frac{{\it J}_{0}}{4} 
  + k \cdot p \; I(k^2)  - p^2 \; I(p^2)  
  +  \frac{1}{2} \, (p^2-k \cdot p) \; I(q^2/4)
   \Bigg\} \; ,    \nonumber  \\
J_{B}(k,p)&=&J_{A}(p,k) \;.  
\end{eqnarray}

\noindent
{\underline {\bf ${\bf J^{(2)}_{\mu\nu}}$}}:
\begin{eqnarray}
{\it J}^{(2)}_{\mu\nu}&=&\frac{{\it i}\pi^2}{2}\Bigg\{
\frac{g_{\mu\nu}}{3}K_0+\left(k_{\mu}k_{\nu}-g_{\mu\nu}\frac{k^2}{3}\right)
J_{C} + 
\left(p_{\mu}k_{\nu}+k_{\mu}p_{\nu}-g_{\mu\nu}\frac{2k\cdot p}{3}\right)
J_{D}  
 + \left(p_{\mu}p_{\nu}-g_{\mu\nu}
\frac{p^2}{3}\right)J_{E} \Bigg\} \;,
\end{eqnarray}
where 
\begin{eqnarray}
J_{C}(k,p)&=&\frac{1}{\Delta^2}   
\Bigg\{ \left[ p^2 (k \cdot p - 2 k^2) - m^2 (k \cdot p - 2 p^2) \right]\; 
\frac{J_A}{2}
 -p^2(p^2-m^2) \; \frac{J_B}{2} \;  \nonumber\\
&& \hspace{10mm} + \frac{k \cdot p}{k^2} \, (m^2-k^2)  \; I(k^2) 
+  \frac{1}{2} \, (k \cdot p + p^2) \; I(q^2/4)  \;
-m \frac{ k\cdot p}{k^2}   \Bigg\} \;,  
   \nonumber \\
J_{D}(k,p)&=&\frac{1}{2\Delta^2}   
\Bigg\{ \hspace{2mm} \left[k^2 (3 k \cdot p - p^2) - m^2 (3 k \cdot p - k^2)
\right] 
\; \frac{J_A}{2} 
 +  \left[p^2 (3 k \cdot p - k^2) - m^2 (3 k \cdot p - p^2)
\right] 
\; \frac{J_B}{2}   \nonumber   \\
&&\hspace{10mm} - (m^2-k^2) \; I(k^2) - (m^2-p^2) I(p^2) - \frac{1}{2} 
(k+p)^2
\; I(q^2/4) + 2m
\Bigg\} \;,  
\nonumber \\
J_{E}(k,p)&=&J_{C}(p,k) \;.
\end{eqnarray}
{\underline {\bf ${\bf I^{(0)}}$}}:
\begin{eqnarray}
   I^{(0)} &=& \frac{1}{\chi} \; \left\{ q^2 (m^2+k\cdot p)
 J^{(0)}  \; + \; i \pi^2 m L  \right\}  \;,
\end{eqnarray}
where
\begin{eqnarray}
  L&=& \frac{q^2(k^2-m^2)-(k^2-p^2)(k^2+m^2)}{(k^2-m^2)^2} \; + \; 
       \frac{q^2(p^2-m^2)+(k^2-p^2)(p^2+m^2)}{(p^2-m^2)^2}  \;.
\end{eqnarray}
{\underline {\bf ${\bf {I^{(1)}_{\mu}}}$}}:
\begin{eqnarray}
{\it I}^{(1)}_{\mu}&=&\frac{{\it i}\pi^2}{2}\left[
 k_{\mu}I_{A}(k,p)+p_{\mu}I_{B}(k,p)\right]  \;,
\end{eqnarray}
where
\begin{eqnarray}
I_{A}(k,p)&=&\frac{2}{\Delta^2}
\Bigg\{  \left[k\cdot p (p^2-m^2) - p^2(k^2-m^2) \right]
\; \frac{I_{0}}{4} + p\cdot q \; \frac{J_{0}}{4} 
+ \frac{mp^2}{(m^2-p^2)^2} -
\frac{mk\cdot p}{(m^2-k^2)^2} \Bigg\} \;,   
\nonumber \\ \nonumber \\
 I_{B}(k,p)&=&I_{A}(p,k)  \;.
\end{eqnarray}

\noindent
{\underline {\bf ${\bf I^{(2)}_{\mu\nu}}$}}:
\begin{eqnarray}
{\it I}^{(2)}_{\mu\nu}&=&\frac{{\it i}\pi^2}{2}\Bigg\{
\frac{g_{\mu\nu}}{3}{\it J}_{0}+\left(k_{\mu}k_{\nu}
-g_{\mu\nu}\frac{k^2}{3}\right)I_{C}
+\left(p_{\mu}k_{\nu}+k_{\mu}p_{\nu}-g_{\mu\nu}\frac{2k\cdot p}{3}\right)I_{D}
+\left(p_{\mu}p_{\nu}-g_{\mu\nu}\frac{p^2}{3}\right)I_{E}
\Bigg\} \;,
\end{eqnarray}
where
\begin{eqnarray}
I_{C}(k,p)&=&\frac{1}{\Delta^2}   \Bigg\{ p^2\;J_{0}  + 
\left[ p^2 (k \cdot p - 2 k^2) -m^2(k\cdot p-2p^2)\right]\; \frac{I_A}{2}
- p^2(p^2-m^2) \; \frac{I_B}{2} \nonumber\\
 && \hspace{9mm} + (k \cdot p - 2 p^2) \; \frac{J_A}{2}  - p^2 \; 
\frac{J_B}{2} \nonumber - \frac{k \cdot p}{k^2 } \; I(k^2)  
+ \frac{mk\cdot p}{k^2(m^2-k^2)}
   \Bigg\}  \;,  
\\  
I_{D}(k,p)&=&\frac{1}{2\Delta^2}   \Bigg\{ - 2 k \cdot p \; J_{0}  +
\left[ k^2 (3 k \cdot p - p^2)-m^2(3k\cdot p-k^2) \right] \; \frac{I_A}{2} 
+  \left[p^2 (3 k \cdot p - k^2)-m^2(3k\cdot p-p^2) \right] 
\; \frac{I_B}{2}   \nonumber \\
&& \hspace{9 mm}
+ (3 k \cdot p - k^2) \; \frac{J_A}{2} 
  +    
  (3 k \cdot p - p^2) \; \frac{J_B}{2} 
\; +\; I(k^2) + I(p^2) \; 
-   \;\frac{m}{m^2-k^2} \; 
- \; \frac{m}{m^2-p^2}  \Bigg\} \,, \nonumber
\\  
I_{E}(k,p)&=&I_{C}(p,k) \;.
\end{eqnarray}
The coefficients $a_{ij}$ in the one loop perturbative expansion of the
$\tau_i$, Eq.~(\ref{taui}),  are tabulated below~:
\begin{eqnarray}
a_{11}(k,p) &=& -(\xi+2)\eta_1(m^2 + k\cdot p)        \nonumber \\
a_{12}(k,p) &=& a_{11}(p,k)                           \nonumber \\
a_{13}(k,p) &=& 4(\xi+2)\frac{(k^2 + k\cdot p)}{(k^2 - p^2)}
                                                      \nonumber \\
a_{14}(k,p) &=& a_{13}(p,k)                           \nonumber \\
a_{15}(k,p) &=& -2(\xi + 2)                           \nonumber \\
a_{16}(k,p) &=& 0                                     \nonumber \\
a_{21}(k,p) &=& -\eta_1\Bigg\{  \left[-\frac{q^{2}}{2}
m^{4}+
\left\{(k \cdot p)^2-(k^{2}+p^{2})(k \cdot p)+k^{2}p^{2} 
\right\} m^{2}-\frac{q^{2}}{4} \left\{ (k \cdot p)^{2}+ k^{2}p^{2}\right\}
\right]                                                \nonumber \\
&+& \frac{(\xi-1)}{2\chi}\Big[-q^{4}  m^{8}- q^2 \left\{
(k\cdot p)^{2}+2(k^{2}+p^{2})k\cdot p-5k^{2}p^{2}\right\}m^{6}
 +\frac{3}{2} q^2 (k^2+p^2) \Delta^2  m^{4}            \nonumber \\
&+& \Big\{2(k^{4}+p^{4}+k^2 p^2)(k\cdot p)^{3}
-7k^{2}p^{2}(k^2+p^2)(k\cdot p)^{2} +10 k^{4}p^{4}k\cdot p
-k^{4}p^{4}(k^2+p^2)\Big\}  m^{2}                      \nonumber \\
&+& \frac{1}{2} k^2 p^2 q^2 \left\{ (k^2+p^2) (k \cdot p)^2 - 
4k^2 p^2 k \cdot p + k^2 p^2 (k^2 + p^2)    \right\}
\Big] \Bigg\}                                          \nonumber \\
a_{22}(k,p) &=& a_{21}(p,k)                           \nonumber \\
a_{23}(k,p) &=& \frac{1}{(k^2 - p^2)}\Bigg[ \xi\left\{(k\cdot p)^{3}+
k^{2}(k\cdot p)^{2}-3k^{2}p^{2}k\cdot p +2k^{4}k\cdot p
+k^{4}p^{2}-2k^{6}\right\}  m^2/k^2                     \nonumber \\
&+&(k\cdot p)^3 +(2k^2-p^2)(k\cdot p)^2+k^2p^2k\cdot p -2k^4 k\cdot p -k^2p^4
                                                        \nonumber \\
&+&(\xi-1)\left\{(k\cdot p)^3+p^2(k\cdot p)^2-3k^2 p^2k\cdot p 
+ 2k^4k\cdot p +k^2p^4-2k^4 p^2\right\} \Bigg]          \nonumber \\
a_{24}(k,p) &=& a_{23}(p,k)                             \nonumber \\
a_{25}(k,p) &=& q^2(m^2 + k\cdot p) + (\xi-1)(q^2 m^2 + (k\cdot p)^2 -
(k^2 + p^2)k\cdot p) + k^2 p^2)                         \nonumber \\
a_{26}(k,p) &=& m \Delta^2\Bigg\{ k\cdot p + \frac{(\xi-1)}{\chi}
\Big[ q^2 k\cdot p m^4 +2(k^2 + p^2)\Delta^2 m^2
 - k^2p^2 \left\{ 2(k\cdot p)^2+(k^2 + p^2)k\cdot p -4 k^2p^2 
\right\}\Big] \Bigg\}                                    \nonumber \\
a_{31}(k,p) &=& -\frac{\eta_1}{2}\Bigg\{\Bigg[\left\{-2(k\cdot p)^{2}
+k^{4}+p^{4}\right\}
 m^{4}                                                   
+2\left\{(k^{2}+p^{2})(k\cdot p)^{2}+(k^2-p^2)^2\,k\cdot p
- k^{2}p^{2}(k^{2}+p^{2})\right\}m^{2}                   \nonumber \\
&+&\frac{1}{2} \left\{ -4 (k \cdot p)^4 +  
(k^{2}+p^{2})^2 (k\cdot p)^{2} + k^{2}p^{2}(k^{2}-p^{2})^2   \right\}\Bigg]
                                                         \nonumber \\
&+&\frac{(\xi-1)}{\chi}\Bigg[q^2
\left\{-2(k\cdot p)^{2}+k^{4}+p^{4}\right\} m^{8}         \nonumber \\
&+& 2 \Big\{ (k^2+p^2) [-2 (k \cdot p)^3 
+ (k^2+p^2) (k \cdot p)^2 + (k^4+p^4) k \cdot p ] 
 - k^2 p^2 (3 k^4 + 3 p^4 -2 k^2 p^2) \Big\} m^{6}        \nonumber \\
&-&\frac{3}{2} q^2 \Delta^2 (k^2-p^2)^2
 m^{4} -2\Big\{ (k^2 + p^2) (k^4+p^4-4k^2p^2) (k \cdot p)^3 
- k^2 p^2 (k^4 + p^4 - 6 k^2 p^2 ) (k \cdot p)^2          \nonumber \\
&+& 2 k^4 p^4 (k^2+p^2) k \cdot p - k^4 p^4 (k^2+p^2)^2  \Big\} m^{2}
-\frac{1}{2} k^2 p^2 q^2 
\Big\{ (k^4+p^4-6k^2 p^2) (k \cdot p)^2
+ k^2 p^2 (k^2 + p^2)^2 \Big\} \Bigg]\Bigg\}              \nonumber \\
a_{32}(k,p) &=& a_{31}(p,k)                              \nonumber \\
a_{33}(k,p) &=&\xi\left\{ (k\cdot p)^{3} -k^{2}(k\cdot p)^{2}-
3k^{2}p^{2}k\cdot p+2k^{4}k\cdot p -k^{4}p^{2}+2k^{6}\right\}
m^2/k^2                                                   \nonumber \\
&+&(\xi-2)\left\{ (k\cdot p)^{3}-(2k^{2}-p^{2})(k\cdot p)^{2}
+k^{2}p^{2}k\cdot p-2k^{4}k\cdot p+k^{2}p^{4}\right\}     \nonumber \\
a_{34}(k,p) &=& a_{33}(k,p)                               \nonumber \\
a_{35}(k,p) &=& -(k^4 + p^4 -2(k\cdot p)^2)\left[\xi m^2 
+(\xi-2)k\cdot p\right]                                   \nonumber \\
a_{36}(k,p) &=& -m\Delta^2\Bigg\{ k\cdot p (k^2 + p^2)+2k^2p^2
+\frac{(\xi-1)}{\chi}\Big[q^2\left\{ (k^2 +p^2)k\cdot p+2k^2 p^2\right\}m^4
+2(k^2 + p^2)^2\Delta^2m^2                                \nonumber \\
&-&k^2p^2(k+p)^2 \left\{ (k^2 + p^2)k\cdot p-2k^2 p^2\right\}\Big]\Bigg\}
                                                          \nonumber \\
a_{41}(k,p) &=& -\eta_1(\xi-1)\frac{(k^2 - p^2)}{2\chi}
\Bigg[ -q^{4}m^{6}+3q^{2} \left\{ -(k^{2}+p^{2}) k\cdot p + 2
k^{2}p^{2} \right\} m^{4}                                  \nonumber \\
 &+&\Big\{  (k\cdot p)^{2} [ 4 (k\cdot p)^{2} - 
3k^4-3p^4-26k^2p^2]  +  k^2 p^2 [
24 (k^{2}+p^{2})  k \cdot p - 3k^4-3p^4-14k^2 p^2)]  \Big\}  
\frac{m^{2}}{2}                                            \nonumber \\
&+&\frac{q^2}{2}
\Big\{ (k^2+p^2) (k \cdot p)^3 + 2 k^2 p^2 (k \cdot p)^2 -
3 k^2 p^2 (k^2 + p^2) k \cdot p + 2 k^4 p^4 \Big \}\Bigg]  \nonumber \\
a_{42}(k,p) &=& -a_{41}(p,k)                                \nonumber \\
a_{43}(k,p) &=& \frac{(\xi-1)}{k^2}\left[ (k^2 + k\cdot p)(k\cdot p)^2 
+k^2(2k^2-3p^2)k\cdot p+k^4(p^2 -2k^2)\right]              \nonumber \\
a_{44}(k,p) &=& - a_{43}(p,k)                              \nonumber \\
a_{45}(k,p) &=& (\xi-1)(k^2-p^2)q^2                        \nonumber \\
a_{46}(k,p) &=& m (\xi-1)(k^2 - p^2)\frac{\Delta^2}{\chi}
\left[ q^2 k\cdot p m^2+2(k^2 + p^2)(k\cdot p)^2
-2k^2 p^2k\cdot p-k^2p^2(k^2+p^2)\right]                   \nonumber \\
a_{51}(k,p)&=&-\eta_1\Bigg\{ \Delta^2 
+\frac{(\xi-1)}{4\chi}\Bigg[- 2 q^4 m^6 
+ 6 q^2\left\{ 2 k^2 p^2 - (k^2+p^2) k \cdot p
\right\} m^4 - 6 k^2 p^2 q^4 m^2                           \nonumber \\
&-& q^2 \left\{ (k^2-p^2)^2 (k \cdot p)^2
+2 k^2 p^2 (k^2 + p^2) k \cdot p - k^2 p^2 (k^2+p^2)^2 \right\} \Bigg]\Bigg\}
                                                            \nonumber \\
a_{52}(k,p) &=& a_{51}(p,k)                                \nonumber \\
a_{53}(k,p) &=& \frac{(\xi-1)}{k^2}\left[(k\cdot p)^2+2k^2k\cdot p
-k^2(2k^2+p^2)\right]                                       \nonumber \\
a_{54}(k,p) &=& a_{53}(p,k)                                 \nonumber \\
a_{55}(k,p) &=& (\xi-1)q^2                                  \nonumber \\
a_{56}(k,p) &=& -m(\xi-1)\frac{\Delta^2}{\chi}
\left[ q^2(k^2+p^2)m^2+2(k^4+ p^4)k\cdot p-2k^2 p^2(k^2+p^2) \right]
                                                            \nonumber \\
a_{61}(k,p) &=& -\eta_1\frac{(k^2-p^2)}{2}\Bigg[ {q^{2}m^{4}}
-2\left\{(k\cdot p)^{2}-(k^{2}+p^{2})k\cdot p+k^{2}p^{2}\right\}m^{2} 
+\frac{q^{2}}{2}\left\{(k\cdot p)^{2}+k^{2}p^{2}\right\}  \nonumber \\
&+&\frac{q^2(\xi-1)}{\chi}\Bigg[q^2 m^8 + 
2 \left\{ (k \cdot p)^2 + (k^2+p^2) k \cdot p - 3 k^2 p^2 \right\} 
m^{6} - \frac{3}{2} q^{2} \Delta^2 m^{4}                    \nonumber \\
&-& 2 \left\{ (k^2+p^2) (k \cdot p)^3 - k^2 p^2 (k \cdot p)^2 
- k^4 p^4\right\}  m^{2} - \frac{1}{2} k^2 p^2 q^2 \left\{ (k \cdot p)^2 
+ k^2 p^2\right\}\Bigg]\Bigg]                               \nonumber \\
a_{62}(k,p) &=& - a_{61}(p,k)                                 \nonumber \\
a_{63}(k,p) &=& -\Bigg[ \xi\left\{ (k^{2}+k\cdot p)(k\cdot p)^{2}
+k^{2}(2k^{2}-3p^{2})k\cdot p-
k^{4}(2k^{2}-p^{2})\right\}m^2/k^2                           \nonumber \\
&+&(\xi-2)\left\{(2k^{2}-p^{2}+k\cdot p) (k\cdot p)^{2} -k^{2}(2k^{2}-p^{2}) 
k\cdot p -k^{2}p^{4}\right\}\Bigg]                           \nonumber \\
a_{64}(k,p) &=& -a_{63}(p,k)                                 \nonumber \\
a_{65}(k,p) &=& -q^2(k^2 -p^2)\left[\xi m^2 -(\xi-2)k\cdot p \right]
                                                             \nonumber \\
a_{66}(k,p) &=& -m(k^2-p^2)\Delta^2\Bigg[ k\cdot p +\frac{(\xi-1)}{\chi}
\left( q^2k\cdot p m^4 +2(k^2 + p^2)\Delta^2 m^2
-k^2 p^2 q^2 k\cdot p\right) \Bigg]                          \nonumber \\
a_{71}(k,p) &=&-\frac{\eta_1(\xi-1)}{4\chi}\Bigg[- 2 q^{6}m^{6}
- 6 q^{4}  \left\{ (k^2+p^2) (k\cdot p) - k^{2} p^{2} \right\} m^4
                                                             \nonumber \\
&-& 3 q^2 \left\{ ((k\cdot p)^2 + k^2 p^2) (k^4+ p^4 + 6 k^2 p^2) 
- 8 k^2 p^2 (k^2+p^2) k\cdot p \right\} m^2                  \nonumber \\
&+&q^2 \Big\{ (k^2-p^2)^2 (k \cdot p)^3 + 
4k^2 p^2 (k^2+p^2) (k \cdot p)^2 -k^2 p^2 (3k^4+3 p^4+10k^2 p^2) k \cdot p + 
4 k^4 p^4 (k^2+p^2) \Big\}\Bigg]                             \nonumber \\
a_{72}(k,p) &=& a_{71}(p,k)                                 \nonumber \\
a_{73}(k,p) &=& (\xi-1)\frac{(k^2-k\cdot p)}{k^2}\left((k\cdot p)^2
+4k^2 k\cdot p -2k^4-3k^2p^2\right)                          \nonumber \\
a_{74}(k,p) &=& a_{73}(p,k)                                  \nonumber \\
a_{75}(k,p) &=& (\xi-1)q^4                                   \nonumber \\
a_{76}(k,p) &=& m(\xi-1)\frac{\Delta^2}{\chi}\Bigg[ q^2
\Bigg\{(k^2+p^2)k\cdot p-2k^2p^2\Bigg\}m^2                   \nonumber \\
&+&2(k^4+p^4)(k\cdot p)^2-4k^2 p^2(k^2+p^2)k\cdot p 
-k^2p^2(k^4+p^4-6k^2p^2)\Bigg]                               \nonumber \\
a_{81}(k,p) &=& -\eta_1\frac{(\xi+2)}{2}q^2(m^2+k\cdot p)    \nonumber \\
a_{82}(k,p) &=& a_{81}(p,k)                                 \nonumber \\
a_{83}(k,p) &=& 2(\xi+2)k\cdot q                             \nonumber \\
a_{84}(k,p) &=& a_{83}(p,k)                                  \nonumber \\
a_{85}(k,p) &=& -(\xi+2)q^2                                  \nonumber \\
a_{86}(k,p) &=& 0    \;.      \label{coeftau}   
\end{eqnarray}

\vfil\eject

{\centerline{\large{\bf{ Figures}}}}

\vspace{0.5cm}
\epsfbox[90 50 -50 250]{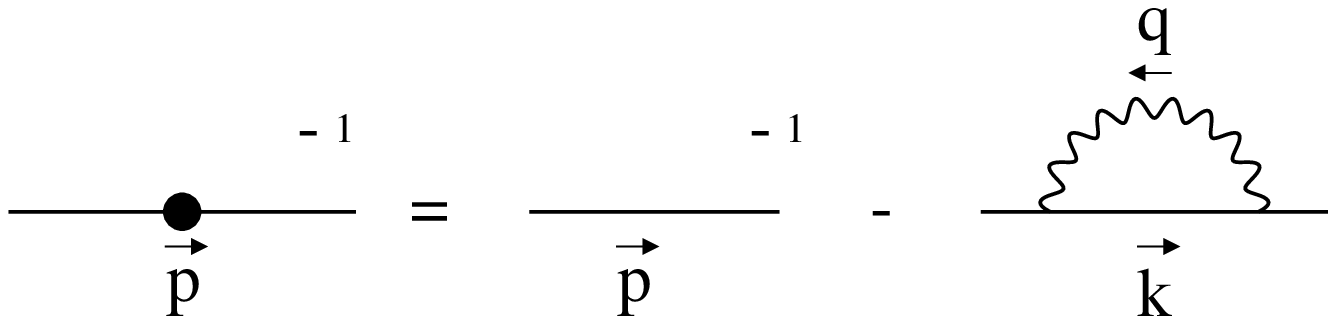}

\vspace{0.5cm}
\noindent
{\hspace{35mm}{\bf Fig.~1.} One loop correction to the fermion propagator.}

\epsfbox[60 50 -125 280]{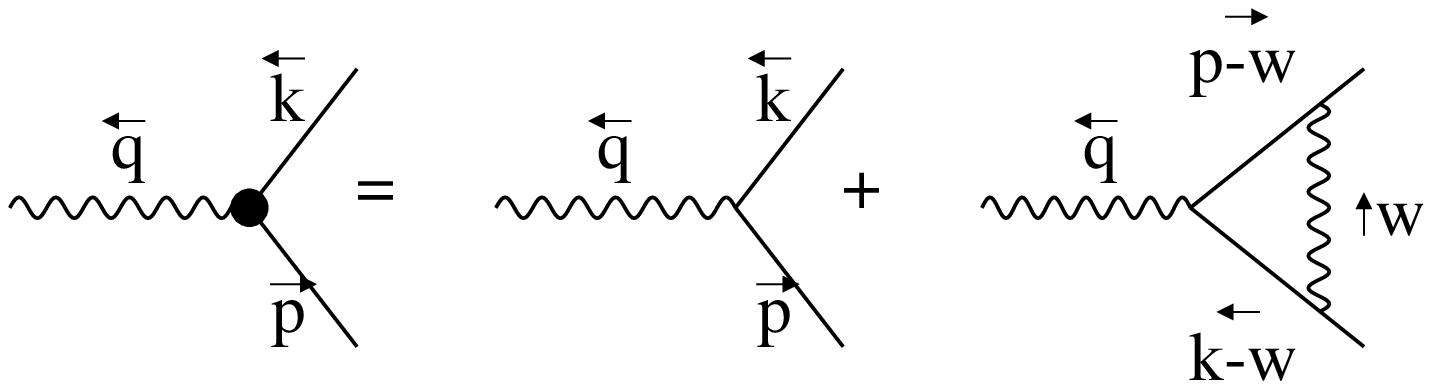}
\vskip 1cm
\noindent
{\hspace{45mm}{\bf Fig.~2.} One loop correction to the vertex.}
\end{document}